\documentclass[showpacs,twocolumn,aps]{revtex4}
\usepackage{amsmath}
\usepackage{times}
\usepackage{epsfig}
\usepackage{amssymb}

\usepackage{graphicx}
\usepackage{epsfig}
\usepackage{psfrag}
\newcommand{\dd}{\text{d}}

\newcommand{\ee}{\text{e}}

\newcommand{\p}{\partial}

\newcommand{\C}{\mathcal{C}}

\begin{document}
\title{Universal cumulants of the current\\
in diffusive systems on a ring} 
\author{C. Appert-Rolland$^{(1)}$, B. Derrida$^{(2)}$,  V. Lecomte$^{(3,4)}$  and F. van Wijland$^{(3)}$}

\affiliation{$^{(1)}$Laboratoire de Physique Th\'eorique
(CNRS UMR 8627), Universit\'e Paris-Sud, b\^atiment 210, 
91405 Orsay, France}

\affiliation{$^{(2)}$Laboratoire de Physique Statistique
(CNRS UMR 8550), \'Ecole Normale Sup\'erieure, 24 rue Lhomond, 75231 Paris cedex 05, France}

\affiliation{$^{(3)}$Laboratoire Mati\`ere et Syst\`emes Complexes 
(CNRS UMR 7057), 10 rue Alice Domon et L\'eonie Duquet, 
Universit\'e Paris VII, 75205 Paris cedex 13, France}

\affiliation{$^{(4)}$D\'epartement de Physique de la Mati\`ere Condens\'ee, Universit\'e de Gen\`eve, 24 quai Ernest Ansermet, 1211 Gen\`eve, Switzerland}

\begin{abstract}
We calculate exactly the first cumulants of the integrated current and of
the activity (which is the total number of changes of configurations) 
of the symmetric simple exclusion process (SSEP)   on a ring with periodic boundary
conditions.  Our  results indicate that for large system sizes the large
deviation functions of the current and of the activity take a universal scaling
form, with the same scaling function for both quantities.
This scaling function can be understood either by an analysis of 
Bethe ansatz equations or in terms of a theory based on
fluctuating hydrodynamics or on the macroscopic fluctuation theory of
Bertini, 
De Sole, Gabrielli, Jona-Lasinio and Landim.
\end{abstract}
\maketitle

\section{Introduction}
The symmetric simple exclusion process (SSEP)  \cite{KOV,HS,Liggett,KL} is one of the simplest lattice gas
models  studied in the theory of non-equilibrium systems. It consists of
hard-core particles hopping with equal rates to either of their nearest
neighbor sites, on a regular lattice. At equilibrium,  when
isolated, the system reaches in the long time limit an equilibrium where all accessible configurations are equally likely. Also, when equilibrium is achieved by contact with one or several reservoirs at a single
density $\rho$, all sites are occupied with this density $\rho$ and
the occupation numbers of different sites are uncorrelated.

As soon as the system is maintained out of equilibrium, by contact with
reservoirs at unequal densities, there is a current  of particles and one observes
long range correlations in the steady state \cite{spohn}.
In this out of equilibrium case several approaches have been developed  to calculate 
steady state properties, such as the  fluctuations or the large
deviations of the density  or of the current \cite{BDGJL1,BDGJL2,BDGJL3,BDGJL5,BDGJL6,DLS1,DLS2,DLS3,ED,bodineauderrida,bodineauderrida2,bodineauderrida3,derrida}.

A lot of progress has been made  over the last decades on
the study of the fluctuations and the large deviation functions  of the current in equilibrium or non equilibrium systems. 
 The large deviation function of the current can be viewed as the dynamical analog of 
 a  free energy, as discussed  by Ruelle in the early seventies~\cite{ruelle}. The idea back then was to build up a thermodynamic formalism based upon probabilities over time realizations rather than over instantaneous configurations. 
 Generic properties of these large deviation functions were later
discovered such as  the fluctuation theorem which determines how the
large deviation function of the current is changed under time reversal
symmetry
~\cite{evanscohenmorriss,gallavotticohen,Kurchan,lebowitzspohn,MRW,Seifert,M1,G2,Gaspard}.

In the present work, we obtain exact expressions for the first cumulants of the integrated current and of the activity (which is the number of changes of configurations) during a long time $t$ for the SSEP consisting of $N$ particles on a ring of $L$ sites. For large system sizes, these cumulants and the associated large deviation functions  take universal scaling forms. We show how these scaling forms can be calculated for the SSEP by the Bethe ansatz or for more general diffusive systems on a ring by   a theory based  on fluctuating hydrodynamics or on the macroscopic fluctuation theory developed  by
Bertini, De Sole, Gabrielli, Jona-Lasinio and Landim~\cite{BDGJL5,BDGJL6,bodineauderrida2,bodineauderrida3,derrida}.
In the Bethe ansatz approach these scaling forms can be extracted from a detailed analysis of finite size effects similar to what was developed recently for quantum spin chains in the context of string theory ~\cite{GK,BTZ}. In the fluctuating hydrodynamics approach, it results from the discreteness of the wave vectors of the fluctuating modes on the ring.

Universal distributions of the current  characteristic of the 
universality class of the KPZ (Kardar-Parisi-Zhang)  equation
\cite{KPZ,PS1,J,RS},  have been calculated in the past 
\cite{derridalebowitz,derridaappert,leekim,OG5} for the  asymmetric
exclusion process (ASEP). The distributions obtained in the present
paper are different and belong to  the Edwards-Wilkinson universality
class \cite{EW}.

We begin by presenting in Sec.\ref{perturbation} exact  expressions of
the first cumulants of the current and of the activity for  the SSEP on a
ring . This is where  we see that the cumulants of the integrated current
and of the activity take scaling forms when the size of the ring becomes
large and where  emerges the idea that the large deviation function of
the current and of the activity obey the same universal scaling function.
This is confirmed in  Sec.\ref{bethe} by   Bethe ansatz calculations. By
resorting to fluctuating hydrodynamics in Sec.\ref{langevin} we are able
to formulate the particular case of the SSEP within a  more general framework  using the Bertini,  De Sole, Gabrielli, Jona-Lasinio and Landim approach  and  to show that the same universal distribution of the current fluctuations of the  current are present in a larger family of diffusive systems.

\section{Exact expressions of the first cumulants}
\label{perturbation}

We consider a system of $N$ particles on a one-dimensional lattice of
$L$ sites with periodic boundary conditions. Each site is either empty
or occupied by a single particle. 
A microscopic  configuration ${\mathcal{C}}=\{n_i\}_{i,1,\ldots,L}$ can
be specified by
$L$ occupation numbers $n_i$ (where $n_i=1$ if site $i$ is
occupied and $n_i=0$ if site $i$ is empty).
In the simple symmetric exclusion
process, SSEP, each particle hops to its right neighbor at rate $1$ or to
its left neighbor at rate 1, provided the target site is empty. 
 In the present paper we
try to determine the distribution of the total integrated current $Q(t)$ and  of the
total number $K(t)$ of changes of configuration (that we will call the activity~\cite{lecomteappertvanwijland}) during a time interval
$(0,t)$. To do so we define the generating functions of the cumulants of $Q$
and $K$ as
\begin{equation}\label{def-psiQK}
  \psi_Q(s)=\lim_{t\to\infty}\frac{\ln\langle\ee^{-s Q}\rangle}{t},\quad
  \psi_K(s)=\lim_{t\to\infty}\frac{\ln\langle\ee^{-s K}\rangle}{t},
\end{equation}
where the brackets denote an average over the time evolutions during
the time interval $(0,t)$. As the evolution is an irreducible  Markov process with a
finite number of states, the long time limits in (\ref{def-psiQK}) do not depend
on the initial configuration and  the generating functions
defined in (\ref{def-psiQK}) can be calculated as the largest eigenvalue of a matrix
\cite{derridalebowitz,lebowitzspohn,derridadoucotroche}. 

Because the calculations are very similar for both observables $K$ and $Q$,
we shall first focus on the activity $K$  and explain 
 how to calculate the cumulant generating function  $\psi_K(s)$ as a
perturbation series in powers of $s$.
We will then present only the results for $\psi_Q(s)$.

\subsection{The cumulants of the activity  $K(t)$}
\label{cumulants-K}
In order to determine $\psi_K$, as in \cite{derridalebowitz}, one can  write
a Master equation for the probability $P(\C,K,t)$ to find the system
in configurations  $\C$ at time $t$, given that the activity at time $t$ is $K$ (i.e.
given  that the system has changed $K$ times of configurations during the time interval $(0,t)$).
\begin{equation}
\p_t P(\C,K,t)\!\!=
\!\!- r(\C) P ({\C},K,t)
\!+\!\sum_{{\C}'} W(\C' \to \C) P ({\C}',K-1,t)
\end{equation}
where $W(\C\to\C')$ is the transition rate from configuration
$\C$ to $\C'$, and  $r(\C)=\sum_{\C'}W(\C\to  \C')$ is the  escape rate 
from configuration $\C$.

If one introduces the generating function 
 $\widehat{P}(\C,s,t)=\sum_{K}\ee^{-sK}P(\C,K,t)$,
its evolution  satisfies
\begin{equation}
\p_t\widehat{P}(\C,s,t)=\sum_{{\C}'}\mathbb{W}_K(\C,{\C}')\widehat{P}({\C}',s,t)
\end{equation}
where
\begin{equation}
\label{WK}
\mathbb{W }_K(\C,\C')=\ee^{-s}W(\C'\to \C)-r(\C)\delta_{\C,\C'} \  .
\end{equation}
In the long time limit,
$\widehat{P}(\C,s,t)$ grows (or decays) exponentially with time, with a rate given by the  eigenvalue  with largest real part \cite{derridalebowitz} 
of
 the modified matrix $ \mathbb{W}_K$. Thus  
$\psi_K(s)$  can be calculated as this largest eigenvalue of  $ \mathbb{W }_K$. 
For $s=0$, $\mathbb{W}_K$ reduces to the
evolution operator of the Master equation $W$ for the symmetric simple exclusion process,
and this  largest eigenvalue (which is  0) as well as  the related
eigenvector are known. We now present a way of obtaining the large deviation
function  $\psi_K$, by  a perturbative expansion \cite{derridadoucotroche,bodineauderridalebowitz}
in powers of $s$.

The idea is to start from the eigenvalue equation for $\psi_K$ and its
eigenvector $\widetilde{P}$, 
\begin{equation}\psi_{K}(s) \widetilde{P}(\C,s) =
 \sum_{\C'} \mathbb{W}_{K}( \C,\C') \widetilde{P}(\C',s)
\end{equation}
normalized such that $\sum_\C  \widetilde{P}(\C,s)= 1$.
One can then define 
  the average  $\langle {\cal A}(\C) \rangle_s$ of an
observable ${\cal A}(\C)$
in the corresponding eigenstate, (i.e.
$\langle {\cal A}(\C) \rangle_s = \sum_\C {\cal A}(\C) \widetilde{P}(\C,s)$
and this is   the same as  averaging, in the  limit of a long time interval $(0,t)$,  
 over all trajectories weighted
by a coefficient $\ee^{-s K(t)}$).
Note that, though the value of $K(t)$ is defined on trajectories running
from $0$ to $t$, the observable ${\cal A}(\C)$  is evaluated at the final time $t$.
From the eigenvalue equation, one gets 
\begin{equation}\label{genobs}
\psi_K(s)\langle {\cal A}(\C) \rangle_s=\ee^{-s} \left\langle \sum_{\C'}W(\C\to\C'){ {\cal A}}(\C') \right\rangle_s-\langle{ {\cal A}}(\C) r(\C) \rangle_s
\end{equation}
where the escape rate  $r(\C)$
is twice the number of clusters of adjacent particles in the system
\begin{equation}
r(\C)= \sum_{\C'}W(\C\to  \C')=2\sum_{j=1}^L n_j(1-n_{j+1}) \  .
\end{equation}
Choosing  ${\cal A}(\C)=1$ in (\ref{genobs})  leads  to
\begin{equation}
\psi_K(s)=(\ee^{-s}-1)\langle r(\C)\rangle_s=2L(\ee^{-s}-1)(\rho-C_s(1))
\label{psik}
\end{equation}
where $C_s(r)=\langle n_i n_{i+r}\rangle_s$ is the correlation function
(which by translational invariance does not depend on $i$) computed within
the eigenstate $\widetilde{P}(\C,s)$, and $\rho=N/L$ is the average density.

For the leading contribution as $s\to 0$, we can use the fact that at  $s=0$ the eigenvector is known (this is the equilibrium distribution,
 for which all allowed microscopic configurations are equally likely), so that
$\psi_K(s)=-2N \left(1-\frac{N-1}{L-1}\right)s+{\cal O}(s^2)$. In order to compute
the ${\cal O}(s^2)$ contribution from (\ref{psik}), we need to  evaluate $C_s(1)$ at order $s$, which can be done 
by choosing ${ {\cal A}}(\C)=n_i n_j$ in  (\ref{genobs}). This requires the knowledge of the correlation function $C_s(r)=\langle n_i n_{i+r}\rangle_s$ at order ${\cal O}(s)$.
For   ${ {\cal A}}(\C)=n_i n_j $ in 
(\ref{genobs}) one gets 
\begin{eqnarray}\nonumber
C_s(1)-C_s(2)= s A_{N,L}+{\cal O}(s^2)\qquad\qquad\qquad\qquad\\\nonumber
{\rm where \ \ \ \ }A_{N,L}=\frac{N(N-1)(L-N)(L-N-1)}{L(L-1)^2(L-2)}\\
C_s(r+1)\!+\!C_s(r-1)\!-\!2 C_s(r)\!= \!s \frac{2  A_{N,L}}{L-3}\!+\!{\cal O}(s^2)\\\nonumber
{\rm for} \ \ \ \ 2 \leq r \leq  L-2,\; 
\end{eqnarray}
which have the following solution
\begin{equation}
C_s(r)=\frac{N(N-1)}{L(L-1)}-sA_{N,L}\frac{6 r (L-r) - L(L+1)}{6(L-3)}+{\cal O}(s^2) \  .
\end{equation}
We can therefore extract $\psi_K$ up to ${\cal O}(s^2)$ and $\langle K^2\rangle_c/t$ follows. 

To obtain  higher  cumulants, we have repeated 
the same procedure,  with the observables
${ {\cal A}}(\C)=n_i n_j n_k$ and 
${ {\cal A}}(\C)=n_i n_j n_k n_l$. The calculations are longer but very
similar.
We found  that the first cumulants of $K$, $\lim_{t \to \infty} \langle K^n\rangle_c/t=(-1)^n\left.\frac{\dd^n\psi_K}{\dd s^n}\right|_{s=0}$, when expressed in terms of the system size $L$ and of 
\begin{equation}
\sigma(\rho)=2\rho(1-\rho) = {2 N (L-N) \over L^2}
\label{sigma}
\end{equation}
  are given by (in the $t\to \infty$ limit)
\begin{widetext}
\begin{eqnarray}\nonumber
\frac{\langle K\rangle}{t}=L^2\frac{\sigma}{L-1},\;\;\;\;\;
\frac{\langle K^2\rangle_c}{t}= \frac{L^2\sigma( L^2\sigma+4L-4)}{6(L-1)^2}\\
\frac{\langle K^3\rangle_c}{t}= \frac{L^2\sigma\big[-L^5\sigma^2+L^4\sigma(2+ 3 \sigma)-2 L^3\sigma+48(L-1)^2\big]}{60(L-1)^3} \nonumber\\
\frac{\langle K^4\rangle_c}{t}=L^2\sigma \Big(\sigma^3L^6(10L^3-70L^2+175L-153)-4\sigma^2L^4(L-1)(11L^3-69L^2+154L-126)\nonumber\\
+16 \sigma L^2(L-1)^2(3L^3-17L^2+46L-63)+2112(L-1)^3(L-3)\Big)\Big({2520(L-1)^4(L-3)}\Big)^{-1} \  .
\label{mom-K}
\end{eqnarray}
\end{widetext}
When $L$ becomes large, while $\rho = N/L$ is kept fixed,
the asymptotic behavior of the above cumulants reads
\begin{eqnarray}
\label{Kasympt}
\frac{\langle K\rangle}{t} \simeq \sigma L,\;\frac{\langle K^2\rangle_c}{t} \simeq
 \frac{\sigma^2}{6} L^2,\\\nonumber
\frac{\langle K^3\rangle_c}{t}\simeq - \frac{\sigma^3}{60}L^4,\;
\frac{\langle K^4\rangle_c}{t}\simeq \frac{\sigma^4}{252}L^6
\end{eqnarray}
One
 might have expected the derivatives at $s=0$ of the eigenvalue
$\psi_K$ to become  extensive for a large system size $L$ (after all, as we shall see
it in section \ref{bethe}, it is always possible to view  $\psi_K$ as the ground state energy of a short range  Hamiltonian). Yet this is  not the case since  the second  and higher cumulants
 grow faster than linearly with $L$ at fixed density $\rho$.
This suggests that, in the large $L$ limit,   $\psi_K/L$
 becomes a  singular function of $s$ at $s=0$.

Also one can guess from (\ref{Kasympt}) that  for $ n\geq 2$ $$\frac{\langle K^n\rangle_c}{t}\sim
\sigma^n
 L^{2 n -2} 
$$ and that for $L\to \infty$ and $s \to 0$, the eigenvalue
$\psi_K$  takes a scaling form 
\begin{eqnarray}\label{scalingpetitarg}
\lim_{L\to\infty}L^2\left[\psi_K(s)+s\frac{\langle K\rangle}{t}\right]={\cal F}_K\left(\frac{\sigma}{2}  L^2s\right)\nonumber\\
\end{eqnarray}
where  the scaling function ${\cal F}_K$ is given by 
\begin{equation}\label{Fexpansionpetitarg}
{\cal F}_K(u)=\frac 13 u^2+\frac{1}{45} u^3+\frac{1}{378}u^4+{\cal O}(u^5) \  .
\end{equation}
We shall see in sections \ref{bethe} and \ref{langevin}  that this scaling function can be fully
determined and  written as
\begin{equation}
 {\cal F}_K \left(  u \right) = -4
 \sum_{n
\geq 1} \left[  
  n \pi \sqrt{n^2 \pi^2 -2 u }  - n^2 \pi^2 + u \right]
\label{F-def}
\end{equation}
or equivalently (see appendix A) as
\begin{equation}
 {\cal F}_K \left(  u \right) = \sum_{k \geq 2}
  { B_{2 k - 2} \over (k-1)!  \ k! } 
(-2u)^k
\label{F-def1}
\end{equation}
where the  Bernoulli numbers $B_n$   are known to be 
simply the coefficients of the expansion  $x (e^x-1)^{-1} = \sum_n B_n x^n
/ n! $.
As a consequence, the generalization of   (\ref{Kasympt})  
will be 
for $n \geq 2$
\begin{equation}
\frac{\langle K^n\rangle_c}{t}\simeq \frac{ B_{2n-2}}{ (n-1)!}  
\sigma^nL^{2 n -2} 
\  .
 \label{Kgen}
\end{equation}

\subsection{The cumulants of the current}

The same procedure can be followed for the total integrated current $Q$
(which can be defined by
$Q= \sum_{j=1}^N x_j(t)$ where $x_j(t)$ is the total displacement of the
$j$th particle during the time interval $(0,t)$).
Its cumulant generating function $\psi_Q$  defined in (\ref{def-psiQK}) is the
eigenvalue (with largest real part) of the  matrix
\begin{equation}
\label{WQ}
\mathbb{W }_Q(\C,\C')=W(\C'\to \C)\ee^{-s j(\C',\C)}-r(\C)\delta_{\C,\C'}
\end{equation}
where $j(\C',\C)$ is $+1$ or $-1$ depending on whether a particle has moved
to the right or to the left when the system jumps from configuration $\cal C'$ to configuration $\cal C$.
Using an expansion in powers of $s$ as in  \ref{cumulants-K} we have obtained
(in the limit $t \to \infty$)
\begin{widetext}
\begin{eqnarray}
\frac{\langle Q^2\rangle}{t}=\frac{L^2\sigma}{L-1},\;\;\;\;\;\;\;\;
\frac{\langle Q^4\rangle_c}{t}=\frac 12 \frac{L^4\sigma^2}{(L-1)^2}
  \\ \nonumber
\frac{\langle Q^6\rangle_c}{t}=
-\frac{ L^6\sigma^2\left((L^2-L+2)\sigma-2(L-1) \right)}{4(L-1)^3(L-2)}\nonumber\\
\frac{\langle Q^8\rangle_c}{t}=
\frac{ L^8\sigma^2\left(   (10L^4-2L^3+27L^2-15L+18)\sigma^2 -4(L-1)(11L^2-L+12)\sigma +48(L-1)^2\right)}{24(L-1)^4(L-2)(L-3)}
\label{mom-Q}
\end{eqnarray}
\end{widetext}
with the corresponding large $L$ behaviors (for $\rho=N/L$ fixed)
\begin{eqnarray} \label{Q-asympt}
\frac{\langle Q^2\rangle_c}{t}\simeq  \sigma  L ,\;
\frac{\langle Q^4\rangle_c}{t}\simeq \frac{\sigma^2}{2} L^2,
 \nonumber \\
\frac{\langle Q^6\rangle_c}{t}\simeq
-\frac{\sigma^3}{4}L^4,\;
\frac{\langle Q^8\rangle_c}{t}\simeq
\frac{5 \sigma^4}{12}  L^6 \  .
\end{eqnarray}
As for $K$,  these results  indicate that  for $ n \geq 2$
$$\frac{\langle Q^{2n}\rangle_c}{t}\sim
\sigma^n L^{2 n-2} $$
and that 
$\psi_Q$ takes  a scaling
form, in the limit $L \to \infty$ and $s \to 0$ 
\begin{eqnarray}\label{scalingpetitarg2}
\lim_{L\to\infty}L^2\left[\psi_Q(s)-\frac{s^2}{2}\frac{\langle Q^2\rangle_c}{t}\right]={\cal F}_Q\left(-\frac  {\sigma}{4} L^2 s^2\right)
\end{eqnarray}
where, according to (\ref{Q-asympt}),  the  expansion of  ${\cal F}_Q(u)$   in powers of $u$ coincides with the expansion 
 (\ref{Fexpansionpetitarg}) of 
 ${\cal F}_K(u)$, at least up to the $4$th order in $u$.

We will see, in section \ref{langevin},  that these  two scaling functions  (which appear 
in (\ref{scalingpetitarg})  and in (\ref{scalingpetitarg2})) are in fact the same.
Therefore the formula which generalizes (\ref{Q-asympt})   
will be 
for $n \geq 2$
\begin{equation}
\frac{\langle Q^{2n}\rangle_c}{t}\simeq { (2n) ! \ B_{2 n -2} \over 2^n
 \ (n-1)!  \ n! } \sigma^n L^{2 n - 2}  \label{Qgen} \  .
\end{equation}
\section{Bethe ansatz}
It is well known that the Bethe ansatz allows one  to calculate the
eigenvalues of
 matrices such as $\mathbb{W }_K(\C,\C') $ and $\mathbb{W }_Q
(\C,\C') $  defined in (\ref{WK},\ref{WQ})  for exclusion processes
\cite{gwaspohn,derridalebowitz,derridaappert,leekim,OG1,Pri,Schutz,OG2,OG4,OG5,OG6,Cantini}.  In this section we show how to obtain the scaling forms (\ref{scalingpetitarg},\ref{scalingpetitarg2}) from the Bethe ansatz equations.

\label{bethe}
\subsection{Relation to spin chains}
It is possible to write the matrices $\mathbb{W }_K(\C,\C') $ and $\mathbb{W }_Q(\C,\C') $  as quantum spin-chain Hamiltonians~\cite{felderhof}. We use the correspondence in which the $z$ component of a two state spin operator is up when a particle is present at site $i$, and is down otherwise. In this basis one finds that
\begin{eqnarray}\label{defHQ-defHK}
\nonumber
\widehat{H}_K=\frac{L}{2}-\frac{1}{2}\sum_{i=1}^L\left[\ee^{-s}(\sigma_i^x\sigma_{i+1}^x+\sigma_i^y\sigma_{i+1}^y)\right.\\
\nonumber
\left.+\sigma_i^z\sigma_{i+1}^z\right] = - \mathbb{W}_K\\
\widehat{H}_Q=\frac{L}{2}-\frac 12
\sum_{i=1}^L\left[\cosh s\, (\sigma_i^x\sigma_{i+1}^x+\sigma_i^y\sigma_{i+1}^y)+\sigma_i^z\sigma_{i+1}^z\right.\\\nonumber\left.-i\sinh s\, (\sigma_i^x\sigma_{i+1}^y-\sigma_i^y\sigma_{i+1}^x)\right]
= - \mathbb{W}_Q
\end{eqnarray}
where we have resorted to the Pauli matrices $\sigma_i^{x,y,z}$. In this language, the quantities $\psi_K$ and $\psi_Q$ are the ground state energies of
these operators. It also suggests that the methods of one-dimensional exactly solvable models apply in our case, such as the Bethe ansatz, as was exploited for similar systems in the past~\cite{kim, schutz2, derridalebowitz}.

As the number of particles on the ring is fixed,   we need to find the
ground state with  a fixed particle density $\rho$, that is, at fixed transverse magnetization $\sum_i\sigma_i^z$.  The quantum operators appearing in (\ref{defHQ-defHK}) have of course been extensively studied~\cite{baxter}, including within the framework of stochastic dynamics~\cite{schutz1}.  For instance, following the notations of Baxter~\cite{baxter} the operator $\ee^{s}\widehat{H}_K$ is the ferromagnetic $XXZ$ chain with anisotropy parameter $\Delta =\ee^s$. Similarly, $\widehat{H}_Q$ corresponds to an $XXZ$ chain with additional Dzyaloshinskii-Moriya interactions.
A study of an operator closely related to $\widehat{H}_Q$ was carried out by Kim~\cite{kim} in 1995. His results will be recalled at the end of the present section.\\

The Bethe ansatz consists in looking for the ground state of $\widehat{H}_{K\text{ or }Q}$ in the form of a linear combination of $N$-particle
plane waves (see \cite{gwaspohn,kim}). We denote by $\{x_j\}_{j=1,\ldots,N}$ the positions of the $N$ particles and we postulate that the right eigenvector of $\mathbb{W}_K$ can be cast in the form
\begin{equation}\label{ansatzwf}
{P}(\{ x_j\},s) = \sum_{\mathcal P}{\mathcal A}({\mathcal P})
\prod_{i=j}^N \left[\zeta_{p(j)}\right]^{x_j}
\end{equation}
where ${\mathcal P}=\left(p(1),\cdots,p(N)\right)$ is a permutation over the first $N$
  integers, and
the $\zeta_j$'s are a priori complex numbers. This is an
exact eigenstate provided these parameters satisfy the so-called Bethe equations. These take different forms for $K$ and $Q$. We now discuss how to implement the Bethe ansatz to calculate $\psi_K(s)$ and $\psi_Q(s)$ defined in (\ref{def-psiQK}). Technical details have been gathered in the appendices.
\subsection{Bethe ansatz for $K$}
For the expression (\ref{ansatzwf}) to be an eigenvector of $\widehat{H}_K$
or $\mathbb{W }_Q$ 
the   $\zeta_j$'s have to satisfy a number of constraints
\cite{yangyang}, the so-called Bethe (see  for example \cite{OG4})
equations
\begin{equation} \label{eqn:Bethe_eqs}
\zeta_i^L=\prod_{\substack{ j=1\\ j\neq i}}^{N} 
\left[-\frac{1-2 \ee^s \zeta_i+\zeta_i \zeta_j}
            {1-2 \ee^s \zeta_j+\zeta_i \zeta_j}\right],
\end{equation}
The expression of $\psi_K(s)$ is given by
\begin{equation} \label{eqn:psi_K_Bethe}
\psi_K(s)=\ee^{-s}\sum_{j=1}^N\left(\zeta_j+\frac{1}{\zeta_j}\right)-2N
\end{equation}

Our goal is to obtain (\ref{scalingpetitarg}) from (\ref{eqn:Bethe_eqs})
and (\ref{eqn:psi_K_Bethe})
in the double limit $s \to 0$ and $ L \to \infty$ keeping $s L^2$ and
$N/L= \rho$ fixed.
Because of the particle-hole symmetry the discussion below is limited to the case $\rho \leq {1 \over 2}$.

In the large $L$ limit,  the $\zeta_j$'s accumulate on a curve which depends
on $s$  and  as $s \to 0^-$ becomes a finite arc of the unit
circle (see \cite{baxter, yangyang} and references therein). Note however that the $s>0$ case can be approached by similar methods.

If one writes 
\begin{equation}
\ee^s = \cos \delta
\label{delta-def}
\end{equation}
 and 
\begin{equation}
 \zeta_j= \ee^{i k_j \delta}
\label{ki-def}
\end{equation}
 (\ref{eqn:Bethe_eqs}) becomes
\begin{equation}
k_i = {1 \over  L } \sum_{\substack{ j=1\\ j\neq i}}^{N}  
U(k_i,k_j) \label{ki}
\end{equation}
where
\begin{equation}U(k_i,k_j) = {1 \over i \delta}  \ln \left[-\frac{1-2   \ee^{i k_i \delta}  \cos \delta + \ee^{i (k_j+k_i) \delta} }
{1-2   \ee^{i k_j \delta}  \cos \delta + \ee^{i (k_j+k_i) \delta} }
\right]\label{UU1}
\end{equation}

In the limit
$\delta \to 0$, one can check that  when $k_i-k_j = {\mathcal
O}(1)$
\begin{equation}
  U(k_i,k_j)
 =  2 \frac{1-k_i k_j}{k_i-k_j}+{\mathcal
O}(\delta^2)  \ .\label{U1} \end{equation} 
In the large  $L$ limit, however, the distance between
consecutive $k_i$ becomes of order $1/L \sim \delta$  and for $i-j$ of
order $1$ one should use instead
\begin{widetext}
\begin{equation}
U(k_i,k_j)= {1 \over i \delta } \ln \left[ \frac{k_i-k_j + i  \delta (1
-  k_i^2) + i \delta k_i (k_i-k_j)  - \delta^2 k_i (1-k_i^2)}  
{k_i-k_j  - i  \delta (1 -  k_i^2)  - i \delta k_i (k_i-k_j)  + \delta^2
k_i (1-k_i^2)}   \right] 
\label{U2}
\end{equation}
Therefore  one can rewrite (\ref{ki}) as
\begin{equation}
 L k_i \simeq   \sum_{\substack{ i- n_0 \leq j \leq i+n_0  \\ j\neq i}} {1 \over i \delta} \ln \left[ \frac{k_i-k_j + i  \delta (1
-  k_i^2) + i \delta k_i (k_i-k_j)  - \delta^2 k_i (1-k_i^2)}
{k_i-k_j  - i  \delta (1 -  k_i^2)  - i \delta k_i (k_i-k_j)  + \delta^2
k_i (1-k_i^2)}   \right] +  \sum_{j \notin [i-n_0,i+n_0]}  2 \frac{1-k_i k_j}{k_i-k_j}
\label{ki1}
\end{equation}
\end{widetext}
where $n_0$ is a  fixed large number  $ 1 \ll n_0 \ll L$, so that one can use 
expression (\ref{U1}) for $|j-i| > n_0$ and (\ref{U2}) for $|j-i|  \leq n_0$.
As shown in appendix B, the two sums  (\ref{sum1},\ref{sum2}) in (\ref{ki1}) depend on the cut-off $n_0$ but this dependence disappears when the two terms in the right hand side of (\ref{ki1}) are added.

In the large $L$ limit, the $k_i$ become dense  on an interval
$(-\theta,\theta)$ of the real axis,   with some
density $ g(k)$. In what follows we
will assume that the $k_i$ are regularly spaced according to this
density, meaning that
\begin{equation}
\label{gk}
L \int_{k_i}^{k_j} g(k) dk = j - i \ \ \ \ \ \ {\rm and}  \ \ \ \ \
L\int_{-\theta}^\theta g(k) dk =N \  .
\end{equation} 
Replacing the two sums in (\ref{ki1})  by their expressions (\ref{sum1},\ref{sum2})
obtained in Appendix B, one gets   that for $k=k_i$ the density $g(k)$ should satisfy
 \begin{equation}\begin{split}
 k=  2 {\mathcal P}\int_{-\theta}^{\theta}\dd k' g(k') {1- k'^2  \over k-k'} 
{1 \over L}
\Bigg[\left(   \frac{ g'(k) (1-k^2) } {g(k) }  - 2 k \right)\\\times 
 \pi (1- k^2) g(k) L \delta  \coth  [\pi (1- k^2) g(k) L \delta ]\Bigg] 
\label{B15}
\end{split}\end{equation}

If we make the change of variable 
 $k'=\theta y$, $k=\theta x$,  and
\begin{equation}
g(k) (1-k^2) = \phi(x) 
\label{g1}
\end{equation}
equation (\ref{B15}) becomes
\begin{equation}
   {\mathcal P}\int_{-1}^{1}\dd y  {\phi(y) \over y-x } =  f(x)
\label{airfoil-solution}
\end{equation}
where
\begin{equation}
f(x)=- { \theta  x \over 2} 
+ {   \pi (1-\theta^2 x^2) \phi'(x)  \over 2   \theta }  \delta 
 \coth \left[L  \delta \pi \phi(x)  \right]  + ...  
\label{f-expression}
\end{equation}

As explained in  (\ref{C1},\ref{C2}) of Appendix C one  can invert (\ref{airfoil-solution}) and  express  $\phi(x)$  in terms of $f(x)$
\begin{equation}
\phi(x)=  {C \over \sqrt{1-x^2}} -
{1 \over \pi^2 \sqrt{1-x^2}} \ {\cal P}\int_{-1}^1 {\sqrt{1-y^2} \over y-x} 
f(y)  dy  
\label{phi-sol}
\end{equation}
where the constant $C$ is so far an arbitrary constant.

For small $\delta$, one can write (\ref{eqn:psi_K_Bethe}), using (\ref{ki-def},\ref{g1},\ref{phi-sol}),  as
\begin{equation}\nonumber\begin{split}
\psi_K(s) \simeq \sum_{j=1}^N \delta^2 (1-k_i^2)  \simeq
L  \delta^2  \int_{-\theta}^\theta g(k) (1-k^2)  dk
\\= L\! \delta^2 \theta\! \left[ \! \int_{-1}^1 \!\!  \!
 dx \!\!{C \over \sqrt{1-x^2}} \! - \!
{1 \over  \pi^2 \sqrt{1-x^2}} \ {\cal P}\!\int_{-1}^1\!\! {\sqrt{1-y^2} \over
y-x}   f(y)  dy  \right]
\end{split}\end{equation}
which gives using (\ref{B3},\ref{B5})
\begin{equation}
\psi_K(s) \simeq L \delta^2  \theta     C \pi 
\label{psig}
\end{equation}


 Also, as  (\ref{gk}) $$\int_{-\theta}^\theta g(k) dk = \rho$$
one has (\ref{g1},\ref{phi-sol})
\begin{equation}\nonumber\begin{split}
\rho = \theta \int_{-1}^1 dx \left[ {C \over (1-\theta^2 x^2) \sqrt{1-x^2} }\right.\\\left. -
{1 \over  \pi^2 (1-\theta^2 x^2) \sqrt{1-x^2} } \ {\cal P} \int_{-1}^1
{\sqrt{1-y^2} \over y-x} f(y) dy \right]
\end{split}\end{equation}
which can be simplified using (\ref{B6},\ref{B8}) 
\begin{equation}
 \rho = {  C \theta  \pi \over \sqrt{1-\theta^2}} + {\theta^3 \over  \pi \sqrt{1-\theta^2}}
\int_{-1}^1 {f(y) y  \sqrt{1-y^2}\over 1 - \theta^2 y^2} dy   \ .
\label{rhog}
\end{equation}

\subsection{The leading order in the large $L$ limit}
For large $L$  (at fixed $L \delta$),
(\ref{f-expression}) reduces to 
 $f(x)= - \theta x/2$, so that (\ref{phi-sol}) becomes to leading order using (\ref{C3bis})
\begin{equation}
\phi(x) = { 4 \pi  C - \theta \over  4  \pi   \sqrt{1-x^2}}+ { \theta \over  2
\pi } \sqrt{1-x^2}  + {\cal O}\left({1 \over L}\right)
\label{phi-sol1}
\end{equation}
whereas (\ref{rhog}) becomes  using (\ref{B8})
\begin{equation}
 \rho  = {  C \theta  \pi \over \sqrt{1-\theta^2}} + 
{1 \over 2} +  {\theta^2 -2\over 4 \sqrt{1-\theta^2}} 
\label{rho1}
\end{equation}

Therefore for a fixed density $\rho$ of particles, the constant $C$ in (\ref{phi-sol},\ref{phi-sol1}) and
the eigenvalue (\ref{psig})  are given, to leading order in
${1 \over L}$, by
\begin{equation}
C= {1 \over \pi \theta} \left[ \left(\rho- {1 \over 2} \right) \sqrt{1 - \theta^2} +
{2 - \theta^2 \over 4} \right] 
\label{C-leading}
\end{equation}
and 
\begin{equation}
\psi_K(s) = L \delta^2 \left[ \left(\rho- {1 \over 2} \right) \sqrt{1 -
\theta^2} +
{2 - \theta^2 \over 4} \right] 
\label{psi-leading}
\end{equation}
So far, the constant $C$ remains undetermined. 

The leading order corresponds to using expression (\ref{U1}) in (\ref{ki}) even when $i$ and $j$ differ by a few units. 
For the continuum description to  be valid,  we are now going to argue
that  $\phi(x)$ should remain finite as $x \to \pm 1 $, or, in terms of the original density $g$, that $g(k)$ remains finite as $k \to \pm \theta$.
This will  impose (see (\ref{phi-sol1})) that $$C=\frac{\theta}{4\pi} \
.$$ Indeed if we order the $N$ solutions $k_i$ and focus  on the ones
closest to $\theta$, $\ldots<k_{N-1}<k_N\leq\theta$, then we may estimate
using (\ref{gk}) the difference  between $k_N$ and $\theta$, or between
$k_{N-1}$ and $k_N$. If $C\neq \frac{\theta}{4\pi}$, then $g(k)\sim
(\theta - k)^{-1/2}$ as $ k \to \theta$ implies  that $k_N- k_{N-1} \sim L^{-2}$ . This 
is not compatible  with  $k_N>\frac{2}{L}\frac{1-\theta^2}{k_N-k_{N-1}}$
(which follows from (\ref{ki},\ref{U1})), where the right hand side of this inequality would be  ${\cal O}(L)$ in contradiction with the fact that $k_N\leq \theta$. Hence we must  have $4\pi C=\theta$, in which case $k_N- k_{N-1} \sim L^{-2/3}$ and  there is no contradiction.

 It then follows that 
\begin{equation}
 \theta=2\sqrt{\rho(1-\rho)}
\label{Q-leading}
\end{equation}
and therefore $\psi_K(s)= L \delta^2 \rho(1-\rho)$ and  (\ref{phi-sol1})
\begin{equation}
\label{phi-leading}
\phi(x) = {\theta  \sqrt{1-x^2} \over 2 \pi} + {\cal O} \left( {1 \over
L} \right)
\end{equation}
\subsection{The next order }
Once  $\phi$ is known to leading order (\ref{phi-leading}), one 
can update the expression (\ref{f-expression})
\begin{equation}
f(x)=- { \theta  x \over 2}
- {    (1-\theta^2 x^2) x  \over 4   \sqrt{1-x^2}  } \delta
 \coth \left[{L  \delta   \theta \sqrt{1-x^2}\over 2}  \right]  + ...
\label{f-expression1}
\end{equation}
and one gets from (\ref{rhog})
\begin{equation}\begin{split}
\label{rhogbis}
 \rho  = {  C \theta  \pi \over \sqrt{1-\theta^2}} + {1  \over 2}
- {\theta^2 -2 \over 4 \sqrt{1-\theta^2}} \\ - {\theta^3 \delta \over  4 \pi \sqrt{1-\theta^2}} \int_{-1}^1 y^2 \coth
  \left[{ L  \delta   \theta \sqrt{1-y^2} \over 2 } \right] dy
\end{split}\end{equation}
Then using the fact that (see (\ref{ap1}) in appendix A)
\begin{equation}
\label{B21}
\int_{-1}^1 y^2 \coth(u \sqrt{1-y^2} ) dy  = 
{\pi \over 2 u}+ {\pi \over 2 u^3} {\cal F} \left( -{u^2 \over 2} \right)
\end{equation}
we get
\begin{equation}\begin{split}
\label{rhogter} 
 \rho  = {  C \theta  \pi \over \sqrt{1-\theta^2}}  
+ {1  \over 2} + {\theta^2 -2 \over 4 \sqrt{1-\theta^2}} 
-{\theta^2 \over 4 L \sqrt{1-\theta^2}}
 \\ - {1 \over  L^3 \delta^2 \sqrt{1-\theta^2}}  
{\cal F} \left( -{L^2 \delta^2 \theta^2
  \over 8} \right)    \end{split} \end{equation}
and this gives  (\ref{psig}) 
\begin{equation}\begin{split}
 \psi_K(s)= L \delta^2 C \pi \theta =  L \delta^2 \left[\left(\rho- {1 \over 2} \right) \sqrt{1-\theta^2} + {2 - \theta^2 \over 4}\right.\\\left. + {\theta^2 \over 4 L}+ {1 \over L^3 \delta^2 }  {\cal F}\left(
-{L^2 \delta^2 \theta^2 \over 8}\right) \right] \label{eq1} \end{split}\end{equation}

The leading order (the first two terms of (\ref{eq1}))  has a minimum
for  $\theta$ given by (\ref{Q-leading}).
Therefore to obtain $\psi_K(s)$  at first order in ${1 \over L}$ one can
simply replace  $\theta$  by  (\ref{Q-leading}) in (\ref{eq1}) and one gets 
\begin{equation}
 \psi_K(s)= { L \delta^2 \theta^2 \over 4} \left(1+ {1 \over L}\right) + {1
\over L^2 }  {\cal F}\left( -{L^2 \delta^2 \theta^2 \over 8}\right) \label{psik1}
\end{equation}
which is equivalent (see  (\ref{delta-def},\ref{Q-leading})) to  (\ref{scalingpetitarg}).
 \\ \ \\ \
It is shown in (\ref{large-u}) of  appendix A that  for large negative $u$
\begin{equation}\label{Flim}
{\cal F}_K(u)\simeq\frac{2^{7/2}}{3\pi}(-u)^{3/2},\;\;u\to-\infty
\end{equation}
This implies that  (\ref{scalingpetitarg}) becomes for small negative $s$ (but large negative $L^2 s$)
\begin{equation}\label{scalinggrandarg1}
\psi_K(s) \simeq L \left[ - 2 s \rho(1- \rho)  + {2^{7 / 2} \over 3 \pi}
\big(- s \rho (1-\rho) \big)^{3 / 2} + ...\right]
\end{equation}
So for $ s$ small, but $L^2 s $ large, the extensivity of $\psi_K(s)$ is
recovered and (\ref{scalinggrandarg1}) gives the beginning of the small $s$ expansion in the large $L$ limit.

One can also notice that the function ${\cal F}(u)$ 
(\ref{F-def}) becomes singular as $u \to {\pi^2 \over 2}$. This indicates
the occurrence of a phase transition  discussed at the end of
section \ref{langevin}: for $u> {\pi^2 \over 2}$ the optimal profile to
reduce $K$ is no longer flat and  the system adopts  a
deformed profile  as in \cite{bodineauderrida2} . In fact in the limit $s \to +\infty$ the configurations which dominate are those formed of a single cluster of particles and the activity is limited to the two boundaries of this cluster.
 
The result (\ref{psik1}) or equivalently (\ref{scalingpetitarg}) with $\cal F$ given by (\ref{F-def})
\begin{equation}
 {\cal F}_K \left(  u \right) = -4
 \sum_{n
\geq 1} \left[
  n \pi \sqrt{n^2 \pi^2 -2 u }  - n^2 \pi^2 + u \right]
\end{equation}
gives the leading finite-size correction to $\psi_K(s)$. These finite corrections have been calculated recently, starting from the Bethe ansatz equations, for several spin chains in the context of string theory   and expressions very similar to our $\cal F$ have been obtained \cite{BTZ}. Note also that a more systematic approach has been developed to calculate the next finite size correction \cite{GK}.

\subsection{Bethe ansatz for $Q$}
The eigenvector corresponding to the largest eigenvalue of
$\mathbb{W }_Q$
 can be written as in (\ref{ansatzwf}), with    the Bethe equations
(\ref{eqn:Bethe_eqs}) replaced by 
\begin{equation}
\zeta_i^L=\prod_{\substack{ j=1\\ j\neq i}}^{N} 
\left[-\frac{\ee^s-2 \zeta_i+\ee^{-s}\zeta_i \zeta_j}
            {\ee^s-2  \zeta_j+\ee^{-s}\zeta_i \zeta_j}\right],
\label{bethe-Q}
\end{equation}

 Given the solutions $\zeta_j$ to (\ref{bethe-Q}), the expression of $\psi_Q$ reads
\begin{equation}
\psi_Q(s)=-2N+e^{-s}  \big[\zeta_1+\ldots+\zeta_N\big]
                +e^{s}\left[\frac{1}{\zeta_1}+\ldots+\frac{1}{\zeta_N}\right]
\end{equation}
By a method following  closely the steps of the Bethe ansatz for $K$, the basic ingredients of which are provided in appendix E, we arrive at the following result for $\psi_Q$,
\begin{equation}\label{finalpsiQ}
\psi_Q(s)=\frac{1}{2}\sigma(\rho)s^2(L+1)+L^{-2}{\cal F}\left(-\frac{L^2s^2\sigma(\rho)}{4}\right)
\end{equation}
which leads to the asymptotic behavior as $L\to\infty$,
\begin{equation}
\frac{\psi_Q(s)}{L}\simeq \frac{1}{2}\sigma(\rho)s^2+\frac{2^{1/2}}{3\pi}\sigma^{3/2}|s|^3
\label{eq2} 
\end{equation}
The Bethe equations (\ref{bethe-Q})  are very close to that considered by
Kim~\cite{kim} who worked out the asymmetric exclusion process case. As
outlined in appendix E, it seems that Kim's results cannot be extended to
the SSEP. We think that this is at the origin of  the discrepancy between
our expression (\ref{eq2})    and what was found earlier (expression
(A.12) of \cite{lebowitzspohn}) for the same quantity $\psi_Q(s)$.

Before concluding this section devoted to the Bethe ansatz, let us  mention that, both for the current or the activity, one can obtain $\psi_Q(s)$ or $\psi_K(s)$ in the $s\to\infty$ limit by directly solving (\ref{bethe-Q}) or (\ref{eqn:Bethe_eqs}). We do not give these expressions here because they are out of the universal regime.

\section{Fluctuating hydrodynamics and the macroscopic fluctuation theory}
\label{langevin}
In this section we are going to show that the expressions (\ref{scalingpetitarg},\ref{scalingpetitarg2}) can be recovered by a macroscopic theory based on hydrodynamical large deviations \cite{KOV,HS,KL}.

\subsection{Calculation of $\psi_Q$ for a general diffusive system and derivation of (\ref{scalingpetitarg2} )}
The macroscopic fluctuation theory developed  by Bertini, De Sole, Gabrielli, Jona-Lasinio and Landim  \cite{BDGJL1,BDGJL2,BDGJL3,BDGJL5,BDGJL6} is based on the fact that,
for a large system of size $L$, the density and the current of a diffusive system take scaling
forms. If one defines $\widehat{\rho}_i(t)$, the density averaged in the
neighborhood of site $i$ at time $t$, and $\widehat{Q}_i(t)$, the total
flux between site $i$ and $i+1$ during time $t$,  these quantities take
 scaling forms \cite{bodineauderrida3,derrida}
\begin{equation}
\widehat{\rho}_i(t) = \rho \left( {i \over L}, {t \over L^2} \right) 
\label{scal1}
\end{equation}
\begin{equation}
\widehat{Q}_i(t) =  L Q  \left( {i \over L}, {t \over L^2} \right) 
\label{scal2}
\end{equation}
This allows one to define a rescaled current $j(x,\tau)$ as $$j(x, \tau)
= { \partial  Q(x,\tau) \over \partial \tau} =  L {d \over dt}
\widehat{Q}_{Lx} \left( { L^2 \tau } \right)  \ .$$
The average microscopic current between site $i$ and $i+1$ is related to the rescaled current $j$ by
$${d \widehat{Q}_i(t) \over dt} =  {1 \over L} \  j \left( {i \over L}, {t \over L^2} \right) $$

From the macroscopic fluctuation
theory  \cite{BDGJL1,BDGJL2,BDGJL3,bodineauderrida3,derrida}, the probability of observing a rescaled current
$j(x,\tau)$ and a density profile $\rho(x,\tau)$ over a time $t=T L^2$ is given by
\begin{equation}\begin{split}
{\rm Pro} (\{\rho(x,\tau), j(x,\tau) \}) \sim \\ \exp \left[-L  \!\int_0^{T }\! d \tau \! \int_0^1\! dx {[j(x,\tau) + D(\rho(x,\tau)) \rho'(x,\tau )]^2 \over 2 \sigma(\rho(x,
\tau))} \right] 
\label{quad1}
\end{split}\end{equation}
where the current $j(x,\tau)$ and the density profile $\rho(x,\tau)$ satisfy the conservation law
\begin{equation}
  {d \rho \over d \tau} = - {d j \over dx} \ .
\label{conserva}
\end{equation}
and the diffusive system under study is characterized by the two functions $D(\rho)$ and $\sigma(\rho)$.
For the SSEP, these functions are known: $D(\rho)=1$ and $\sigma(\rho)=2 \rho(1-\rho)$  (see  \cite{HS}).

Note that (\ref{quad1}) can be seen as the fact that the macroscopic
density $\rho(x,\tau)$ and the macroscopic current $j(x,\tau)$ satisfy in
addition to the conservation law (\ref{conserva}) a Langevin  equation of
the form  \cite{spohn}.
\begin{equation}
j(x,\tau)=-\p_x \rho(x,t) +\xi(x,\tau)
\end{equation}
 where $\xi(x,\tau)$ is a  Gaussian white noise
\begin{equation}\label{defEW}
\langle\xi(x,\tau)\xi(x',\tau')\rangle=L^{-1}\sigma(\rho(x,\tau))\delta(x-x')\delta(\tau-\tau') \  .
\end{equation}

The contribution of a small time dependent perturbation to a constant
profile $ \rho_0$ and a constant rescaled current $j_0$,
$$\rho(x, \tau) = \rho_0 + \delta \rho(x, \tau)$$
$$j(x, \tau) = j_0 + \delta  j(x, \tau)$$
 to the quadratic form  in (\ref{quad1})  is  
\begin{equation}\begin{split}
 {[j(x,t) + D(\rho(x,t)) \rho'(x,t)]^2 \over 2 \sigma(\rho(x,t))}
  =   {j_0^2 \over 2 \sigma} + {j_0 \over  \sigma} \delta j  
-{j_0^2  \sigma' \over 2 \sigma^2}\delta \rho 
\\+ {j_0 D \over \sigma} \delta \rho' 
 + { \delta j^2 + 2 D \delta j \delta \rho' + D^2 \delta \rho'^2 + 2 j_0 D' \delta \rho \delta \rho' \over 2\sigma}
\\  -{j_0 \sigma' (   \delta j \delta \rho +   D \delta \rho \delta
\rho') \over  \sigma^2} + j_0^2 \left( {\sigma'^2  \over 2 \sigma^3} - {\sigma'' \over 4 \sigma^2} \right) \delta \rho^2 
\label{quad2}
\end{split}\end{equation}
where the functions $D, \sigma,\sigma,\sigma''$ are evaluated at the density $\rho_0$.

If one considers a fluctuation of the form
\begin{equation}
\delta \rho = k[ a_{k,\omega} e^{i \omega \tau + i k x} + a_{k,\omega}^*
e^{-i \omega t - i k x}] 
\label{fluctuation}
\end{equation}
 one has
$$\delta \rho' = ik^2[ a_{k,\omega} e^{i \omega \tau + i k x} -
a_{k,\omega}^* e^{-i \omega t - i k x}] $$
and due to (\ref{conserva})
$$\delta j = - \omega[ a_{k,\omega} e^{i \omega \tau + i k x} +
a_{k,\omega}^* e^{-i \omega t - i k x}]  \ . $$
The ring geometry ($x \equiv x+1$) imposes that the wave numbers $k$ are
discrete
 $$k= 2 \pi n \ \ \ \ {\rm with }    \ \ \  n \geq 1 $$
Also because one considers a finite time interval $T$, the frequencies
$\omega$ are also discrete
and  $$\omega= {2 \pi m   \over T} \ \ \ \ {\rm with }    \ \ \  m  \in
\mathbb{Z} $$

Integrating over the time interval $0 < \tau < T$ and over space, one
gets
one has
\[ \langle \delta \rho^2 \rangle  = 2 k^2 |a_{k,\omega}|^2 T  \]
\[ \langle \delta \rho'^2 \rangle  = 2 k^4 |a_{k,\omega}|^2 T  \]
\[ \langle \delta j^2 \rangle  = 2 \omega^2 |a_{k,\omega}|^2 T  \]
\[ \langle \delta j \delta \rho  \rangle  = -2 k \omega |a_{k,\omega}|^2
T \]
\[ \langle \delta \rho \delta \rho' \rangle  = \langle \delta j
\delta \rho' \rangle  =  0 \]
Therefore the superposition of all the fluctuations (\ref{fluctuation}) leads to
\begin{equation}\begin{split}\nonumber
{\rm Pro}(j_0,\{a_{k,\omega}\}) \sim 
\exp \left[- {j_0^2 \over  2\sigma} {t \over L} 
\right.\\
\left.- {t \over L} \sum_{\omega,k} |a_{k,\omega}|^2 \left(  {(\sigma \omega +
  j_0 \sigma' k)^2 \over \sigma^3} + {D^2 k^4 \over \sigma} - {j_0^2
\sigma'' k^2 \over 2 \sigma^2} \right) \right]
\end{split}\end{equation}
where some terms independent of $j_0$ have been forgotten (they will
be fixed later by normalization).
After integrating over the Gaussian fluctuations and if one replaces the sum
over $\omega$ by an integral one gets
\begin{equation}\begin{split}
{\rm Pro}(j_0) \sim \exp \left[- {j_0^2 \over 2 \sigma} {t \over L} \qquad\qquad
\right.\\
\left.- {t  \over 2 \pi L^2}\!\!\!\! \!\!\sum_{ 1 \leq k \leq k_{\max} }
 \!\! \!\!\int_{-\omega_{\max}}^{\omega_{\max}}\!\! \!\! \!\!\!d \omega
\!  \ln  \left( \! {( \omega  \sigma + j_0 \sigma' k)^2 \over \sigma^3}\! +\!
{D^2 k^4 \over \sigma}\! - \!{j_0^2 \sigma'' k^2 \over 2 \sigma^2}\! \right)\! \right]
\label{quad2a}
\end{split}\end{equation}
where we have introduced cut-offs $k_{\max}$ and $\omega_{\max}$.
The reason for these cut-offs is that the macroscopic fluctuation theory
(\ref{quad1}) is
valid only on hydrodynamic space and time scales. For $x= O(L^{-1})$ or
$\tau = O(L^{-2})$
it  has no validity at all, 
meaning that the cut-offs should satisfy  $k_{\max} < L$ and
$\omega_{\max} < L^2$. 

For large $L$, i.e. for large $k_{\max}$ and $\omega_{\max}$, one can see
by integrating over $\omega$ that only the constant term and the
term proportional to $j_0^2$  depend on the cut-offs so that

\begin{equation}\begin{split}
\label{quad3}
{1 \over 2 \pi}\! \!\! \!\sum_{ 1 \leq k \leq k_{\max} }
 \! \!\! \! \int_{-\omega_{\max}}^{\omega_{\max}} \! \!\! \! d \omega 
  \ln  \left(  {( \omega  \sigma + j_0 \sigma' k)^2 \over \sigma^3} +
{D^2 k^4 \over \sigma} - {j_0^2 \sigma'' k^2 \over 2 \sigma^2} \right) 
\\
\simeq A(k_{\max}, \omega_{\max}) + B(k_{\max}, \omega_{\max})  j_0^2\qquad\qquad\qquad\qquad\\
+  
\sum_{n =1 }^\infty \left\{ \sqrt{ D^2 (2 \pi n)^4 - {j_0^2
  \sigma''\over 2\sigma} (2 \pi n)^2} -  4 \pi^2 n^2 D + {j_0^2
  \sigma''\over 4 D\sigma}
\right\}
\\  
= A(k_{\max}, \omega_{\max}) + B(k_{\max}, \omega_{\max})  j_0^2 
-D {\cal F} \left( {j_0^2  \sigma''\over 16 D^2 \sigma} \right)
\end{split}\end{equation}
where we have used the definition (\ref{F-def}) of $\cal F$.

If the averaged rescaled current is $j_0$ over a macroscopic time $T$, the sum of the microscopic flux over all the bonds is $Q= T L^2 j_0= t j_0$.
Thus as  $\lim_{t \to \infty} {\langle Q^2 \rangle \over t} =
{L^2 \over L-1} \sigma$ (see (\ref{mom-Q}))
one  can determine the cut-off dependent constants and  get
\begin{equation}
{\rm Pro}(j_0) \sim \exp \left[- {j_0^2 (L-1)\over 2 \sigma L^2} {t \over L} 
+ {t \over   L^2}
  D {\cal F} \left({ j_0^2 \sigma''\over 16 D^2\sigma} \right) 
\right]
\label{quad4}
\end{equation}
 where $\cal F$ is defined in (\ref{F-def}).
This becomes, 
at order $1/L^2$,
using the fact that $\psi_Q(s) = \max_{j_0} [ -j0 s +
t^{-1} \ln  {\rm Pro}(j_0)] $ 
\begin{equation}
\psi_Q(s) - {s^2 \langle Q^2 \rangle \over 2 t} = {1 \over L^2}  D {\cal F}
\left({ \sigma \sigma''  \over 16  D^2} L^2 s^2 \right) \  .
\label{psis}
\end{equation}
This formula is in principle valid for arbitrary diffusive systems, i.e.
for arbitrary functions $\sigma(\rho)$ and $D(\rho)$.  As
 $\sigma= 2 \rho (1-\rho)$, $D=1$, $\sigma''=-4$  for the SSEP  this leads to
the announced result (\ref{scalingpetitarg2},\ref{F-def}).

For a general diffusive system  the expressions of the cumulants (\ref{Q-asympt}) would therefore become 
\begin{equation}
\lim_{t \to \infty} {\langle Q^{2  n} \rangle_c \over t} = B_{2 n -2} {(2n)! \over n! \ (n-1)!} D \left({- \sigma \sigma'' \over 8 D^2 } \right)^n L^{2 n - 2} 
\label{cumulants-general}
\end{equation}
where $\sigma(\rho)$ and $D(\rho)$ are the two functions which appear in (\ref{quad1}) and the $B_n$'s are the Bernoulli numbers.

\subsection{Calculation of $\psi_K$ for the SSEP and derivation of (\ref{scalingpetitarg} )}
To obtain (\ref{scalingpetitarg}),
one can first write the activity $K$ as
$$K = 2 L^3 \int_0^T d \tau \int_0^1 dx  \rho(x,\tau) (1 -
\rho(x,\tau)). $$
Then one has
$$K - \langle K \rangle \simeq  2 L^3 \int_0^T d \tau \int_0^1 dx  [  \langle
\delta \rho^2 \rangle - \delta \rho(x,\tau)^2  ]$$

Then one can proceed as above (\ref{quad2}-\ref{psis})  and get, up to terms constant or proportional
to $s$, in the exponential
\begin{equation}\nonumber\begin{split}
\langle e^{-s (K - \langle K \rangle)} \rangle \sim \int d j_0 \int d a_{k,\omega}
\exp \left[- {j_0^2 \over  2\sigma} {t \over L}\right.
\\\left.- \!{t \over L} \!\sum_{\omega,k} \!|a_{k,\omega}|^2 \!\left( \! {(\sigma \omega +
  j_0 \sigma' k)^2 \over \sigma^3} \!+ \!{D^2 k^4 \over \sigma}\! - \!{j_0^2
\sigma'' k^2 \over 2 \sigma^2} \!+ \!4 k^2 s L^2 \right)\! \right]
\end{split}\end{equation}
The rest of the calculation is the same as (\ref{quad3}-\ref{psis}), with a maximum over $j_0$
achieved at $j_0=0$,
and one finally gets
\begin{equation}
\psi_K(s) 
=-s\frac{\langle K\rangle}{t}+L^{-2}{\cal F}_K\left(\frac {\sigma}
{2} L^2s\right)
\label{psikf}
\end{equation}
which is exactly 
 (\ref{scalingpetitarg}).

\subsection{Calculation of $\psi_Q$ in the case of  a weak asymmetry}
One can also repeat the above calculation in the case of weakly driven
systems, i.e. for systems where there is an additional driving force of
strength $1/L$. This would in particular be the case for the weakly asymmetric
exclusion process (WASEP) \cite{bodineauderrida2}  for which the hopping rates to the right and to
the left are respectively $\exp{\nu \over L} $  and $\exp (-{\nu \over
L}) $.

For such systems, 
(\ref{quad1}) becomes
\begin{equation}\begin{split}
{\rm Pro} (\{\rho(x,\tau), j(x,\tau) \}) \sim \\ \exp \left[-L  \!\!\int_0^{T }
\!\! \! d \tau\!\! \!\! \int_0^1 \!\! \!dx {[j(x,\tau) \!+ \!D(\rho(x,\tau)) \rho'(x,\tau )\!-\! \nu
\sigma(\rho(x,\tau)]^2 \over
2 \sigma(\rho(x,\tau))} \right]
\label{quad1a}
\end{split}\end{equation}          
Following exactly the same steps as before, one gets an additional term
${\nu^2 \sigma'' \over 4} \delta \rho^2$ in (\ref{quad2}),  everything
else remaining the same. Then
(\ref{quad4}) becomes in this case:
\begin{equation}\begin{split}
{\rm Pro}(j_0) \sim \exp \left[- {(j_0- \nu \sigma)^2 (L-1)\over 2 \sigma L^2} {t \over
L}\right.
\\\left.+ {t \over   L^2}
  D {\cal F} \left({  
(j_0^2 - \nu^2 \sigma^2) \sigma''
\over 16 D^2\sigma }
 \right)
\right]
\label{quad4a}
\end{split}\end{equation}
where we have adjusted  as in (\ref{quad4}) the terms linear  and
quadratic in $j_0$ which are cut-off dependent.

\subsection{Phase transitions}
The function ${\cal F}(u)$ becomes singular as
$
u \to {\pi^2 \over 2}$ (see (\ref{F-def})).  For systems for which $\sigma'' <0$, this implies the occurrence of a phase transition in  the expression (\ref{psikf}) of  $\psi_K(s)$   in
or in the large deviation function (\ref{quad4a}) of the current in the
case of a weak asymmetry. These phase transitions are exactly the same as
the one discussed in \cite{bodineauderrida2,BDGJL5,BDGJL6,bodineauderrida3}: beyond the transition the
system does not fluctuate anymore about a flat density profile, but the
 profile becomes deformed on a macroscopic scale.

For  systems such as the  Kipnis Marchioro
Presutti  model \cite{KMP,BGL} which have $\sigma''>0$, a similar phase transition 
occurs in $\psi_Q$ even in absence of a  weak asymmetry.

\section{Conclusion}\label{conclusion}
In the present paper we have obtained exact expressions (\ref{mom-K},\ref{mom-Q})  of the first cumulants of the activity $K$ and of the integrated current $Q$ for the SSEP.
In the large $L$ limit, these cumulants take scaling forms (\ref{Kasympt},\ref{Q-asympt}).

We have shown in section \ref{bethe} that these scaling forms can be understood starting from the Bethe ansatz equations (\ref{eqn:Bethe_eqs},\ref{bethe-Q}), by calculating the leading finite size corrections. These finite size corrections are similar to the ones calculated recently for spin chains in the context of quantum strings \cite{BTZ,GK}.

We have also shown  in section \ref{langevin} that they can also be understood starting from the macroscopic fluctuation theory (\ref{quad1})  of Bertini,
De Sole, Gabrielli, Jona-Lasinio and Landim. This enabled us to extend (\ref{psis},\ref{quad4},\ref{cumulants-general}) our results for the SSEP to arbitrary diffusive systems and to see that the occurrence of phase transitions can be predicted from the scaling form of the cumulants of the current.  In order to better understand these phase transitions it might be interesting to characterize the eigenstate of the $s$-dependent evolution operator by, {\it e.g.}, determining correlation functions in those states.\\

 We have discussed here systems governed by diffusive dynamics with a
single conserved field.  How the universal scaling forms would be   modified for systems with several conserved fields is an interesting open question.

\bigskip

We thank  N. Gromov, H.J. Hilhorst,  V. Kazakov,  K.
Mallick, S. Prohlac, H. Spohn, P. Vieira, R.K.P. Zia,  for several useful discussions.  This work was supported
by the French Ministry of Education through an ANR-05-JCJC-44482 grant and  LHMSHE.


\begin{center}
APPENDIX A: SEVERAL REPRESENTATIONS OF THE FUNCTION  $\cal F$
\end{center}
In this appendix we show the equivalence between several representations (\ref{F-def}, \ref{F-def1},\ref{B21})
of the function $\cal F$ defined in (\ref{F-def})
\begin{equation}
\label{f-def-1}
 {\cal F} \left(  u \right) = -4
 \sum_{n
\geq 1} \left[
  n \pi \sqrt{n^2 \pi^2 -2 u }  - n^2 \pi^2 + u \right]
\end{equation}
To do so consider the integral $I$
$$I={2 u^3 \over \pi}\int_{-1}^1 y^2 dy \coth(u \sqrt{1-y^2})$$
Then by
using the fact that
$$\coth z = {1 \over z} + \sum_{n=1}^\infty {2 z \over z^2 + n^2 \pi
^2}$$
and by integrating over $y$, one gets 
\begin{equation}\begin{split}
I=& {2 u^3 \over \pi} \int_{-1}^1 y^2 dy \coth(u \sqrt{1-y^2}) \\
 =&  u^2 + \sum_{n \geq 1} \left[  2 u^2 
 +    4 n^2 \pi^2 -   4  n \pi \sqrt{n^2 \pi^2 + u^2 } \right]\\
=&u^2 + {\cal F} \left( - {u^2 \over 2 } \right) 
\label{ap1}
\end{split}\end{equation}
This establishes (\ref{B21}).
Now as
\begin{equation}
{x \over e^x - 1} = \sum_{n \geq 0} {B_n \over n!} x^n = 1 - {x \over 2} + {x^2 \over 12} - {x^4 \over 720} + {x^6 \over 30240} + ... 
\label{bernoulli}
\end{equation}
 which is
simply the definition of the  
Bernoulli numbers $B_n$ (so that $B_2={1 \over 6}, B_4 = -{1 \over 30},
B_6 = {1 \over 42},...$), one can show that
\begin{equation}
\coth x = {1 \over x} + \sum_{k \geq 2} 2^{2k-2} x^{2 k-3} {B_{2k-2}
\over (2k-2)!}
\label{Bernoulli}
\end{equation} 
Therefore
\begin{equation}\nonumber\begin{split}
I= 
{2 u^3 \over \pi}\int_{-1}^1 y^2 dy \coth(u \sqrt{1-y^2})
\\= {2 u^2 \over \pi} \int_{-1}^1 {y^2 \over \sqrt{1-y^2}} dy
\\+ \sum_{k \geq 2} {2^{2k-1} \over \pi} {B_{2 k-2} \over (2k-2)!} u^{2k} \int_{-1}^1
y^2 (1-y^2)^{2k-3 \over 2} dy 
\end{split}\end{equation}
i.e.
\begin{equation}
I=
  u^2 +
 \sum_{k \geq 2}{B_{2 k-2} \over \Gamma(k) \Gamma(k+1)} u^{2k} 
\label{ap2}
\end{equation}
Comparing (\ref{ap1}) and (\ref{ap2}), one gets
\begin{equation}
 {\cal F} \left(  u \right) =
 \sum_{k \geq 2}{B_{2 k-2} \over \Gamma(k) \Gamma(k+1)} (-2u)^{k} 
  =
 {u^2 \over 3} +{ u^3 \over 45} +  {u^4 \over 378} +
{u^5 \over 2700} + ...
\end{equation}
so that (\ref{F-def1}) and (\ref{Fexpansionpetitarg}) are consistent with
(\ref{F-def}).

For large negative $u$, one gets, by replacing 
in  (\ref{f-def-1}) 
the sum over $n$ by an
integral, 
\begin{equation}
\label{large-u}
{\cal F} \left(  u \right) \simeq {2^{7/2} (-u)^{3/2} \over 3 \pi }  \  .
\end{equation}

\begin{center}
APPENDIX B: CALCULATION OF THE TWO SUMS APPEARING IN (\ref{ki1})
\end{center}
In this appendix we calculate the two sums which appear in (\ref{ki1}) when $\delta \to 0$ and $ L \to \infty$ keeping $L \delta $ fixed.
\\ \ \\
{\bf The first sum in (\ref{ki1})}: \\
If the $k_i$ are distributed according to a density $g(k)$ on the real axis one can write that
\begin{equation}
  L \int_{k_i}^{k_{i+n}} g(k') dk' = n 
\label{gk1}
\end{equation}
Therefore for $n$ fixed and large $L$, one has
$$ L (k_{i+n}-k_i) g(k_i) + L (k_{i+n}-k_i)^2 \frac{g'( k_i) }{2} + ... = n $$
so that
\begin{equation}
k_{i+n}-k_i = \frac{n}{ g(k_i) L }  - \frac{n^2 g'(k_i) }{2  g(k_i)^3 L^2}
+ ...
\label{kin}
\end{equation}
Replacing $k_j$ by expression (\ref{kin}) into the first sum in (\ref{ki1}) one gets 
\begin{equation}\nonumber\begin{split}
\sum_{j=i-n_0}^{i-1}+\sum_{j=i+1}^{i+n_0} U(k_i,k_j) \simeq
\\\sum_{n=1}^{n_0} 
\left( 4 k_i  -\frac{2 g'(k_i) (1-k_i^2) } {g(k_i) } \right)  \frac{n^2 } { n^2
   +  (1 - k_i^2)^2 g(k_i)^2L^2 \delta^2} \ .\end{split}\end{equation}
Using the fact that for  $n_0 \gg 1 $ (and  $b < {\mathcal O}(1)$)
\begin{equation} 
\label{sum}
\sum_{n=1}^{n_0} {1 \over n^2 + b^2} 
=-{1 \over 2 b^2} + {\pi \over 2 b} \coth \pi b 
\end{equation} 
 the first sum in (\ref{ki1}) can be replaced by
\begin{equation}\begin{split}
\sum_{j=i-n_0}^{i-1}+\sum_{j=i+1}^{i+n_0} U(k_i,k_j) \simeq
\left( 4  k_i - \frac{2 g'(k_i) (1-k_i^2) } {g(k_i) }  \right) n_0\\
  - \left( 2 k_i  - \frac{ g'(k_i) (1-k_i^2) } {g(k_i) } \right)  
 \Big[ -1 \qquad\qquad\qquad\\+ \pi (1-k_i^2) g(k_i) L \delta \coth[ \pi (1-k_i^2) g(k_i) L
\delta] \Big]
\label{sum1}
\end{split}\end{equation}

\bigskip
\ \\
{\bf The second sum in (\ref{ki1})}: \\
Let us consider the following integral.
\begin{equation}
I={\mathcal P}\int_{-\theta}^{\theta} g(k') \dd k'  {1- k_i k' \over k_i-k'}
\label{Idef}
\end{equation}
We are now going to  compare this integral with the sum
$$S
=\sum_{j \notin[i-n_0,i+n_0]}   {1- k_i k_j \over k_i-k_j}$$
We assume (\ref{gk1}) that  the $k_j $ are given by
\begin{equation}
L \int_{-\theta}^{k_j} g(q) dq = j - \alpha
\label{jalpha}
\end{equation}
and for the moment $\alpha $ is arbitrary. Therefore 
\begin{equation}
k_{j+1} - k_j \simeq {1 \over g(k_i) L} 
\label{jalpha-bis}
\end{equation}
One can decompose the integral $I$ as
\begin{equation}\begin{split}
I={\mathcal P}\int_{k_{i-n_0}}^{k_{i+n_0}} g(q) \dd q  {1- k_i q \over k_i-q}
\\+\!\!\! \sum_{j=1}^{i-n_0-1}\!\!\! \int_{k_j}^{k_{j+1}} \!\!\! g(q) \dd q  {1- k_i q \over k_i-q}
+\!\!\! \sum_{j=i+n_0}^{N-1}\!\!\!  \int_{k_j}^{k_{j+1}}\!\!\!  g(q) \dd q  {1- k_i q \over k_i-q}
 \
 \\ 
+ \int_{-\theta}^{k_1} \!\!\! g(q) \dd q  {1- k_i q \over k_i-q}
+ \int_{k_N}^{\theta} \!\!\! g(q) \dd q  {1- k_i q \over k_i-q}
\end{split}
\label{X0}
\end{equation}
 As $k_{j+1} -k_j$ is small and of order $1/L$ and because of (\ref{jalpha},\ref{jalpha-bis})
\begin{eqnarray}
\nonumber
\int_{k_j}^{k_{j+1}} g(q) \dd q  {1- k_i q \over k_i-q} \simeq 
{1 \over L} {1- k_i k_j \over k_i-k_j} \qquad\qquad
\\\nonumber+  {g(k_j)(k_{j+1}-k_j)^2 \over 2} \  {d \over dk_j} \left({1- k_i k_j \over k_i-k_j} \right)
\\ \nonumber 
\simeq{1 \over L} {1- k_i k_j \over k_i-k_j} 
+ {1 \over 2 L^2 g(k_j)} \  {d \over dk_j} \left({1- k_i k_j \over k_i-k_j} \right)
\\ \label{X2}
\simeq{1 \over L} {1- k_i k_{j+1} \over k_i-k_{j+1}} 
- {1 \over 2 L^2 g(k_{j+1})} \  {d \over dk_{j+1}} \left({1- k_i k_{j+1} \over k_i-k_{j+1}} \right)
\end{eqnarray}

Therefore using (\ref{X2}) in the sum $1 \leq j \leq i-n_0-1$ and (\ref{X2}) in the sum $i+n_0 \leq j \leq N-1$, one can rewrite (\ref{X0}) as

\begin{eqnarray}
\nonumber
I \simeq {\mathcal P}\int_{k_{i-n_0}}^{k_{i+n_0}} g(q) \dd q  {1- k_i q \over k_i-q}
+{1 \over L} \sum_{j=1}^{i-n_0-1}  {1- k_i k_j \over k_i-k_j}
\\\nonumber+{1 \over L} \sum_{j=i+n_0+1}^{N}  {1- k_i k_j \over k_i-k_j}
+{1 \over 2 L^2} \sum_{j=1}^{i-n_0-1}  
{1 \over g(k_j)} \ {d \over dk_j} \left({1- k_i k_j \over k_i-k_j} \right)
\\\nonumber-{1 \over 2 L^2} \sum_{j=i+n_0+1}^{N} 
{1 \over g(k_j)} \ {d \over dk_j} \left({1- k_i k_j \over k_i-k_j} \right)
\\\nonumber+ \int_{-\theta}^{k_1} g(q) \dd q  {1- k_i q \over k_i-q}
+ \int_{k_N}^{\theta} g(q) \dd q  {1- k_i q \over k_i-q}
\end{eqnarray}
This becomes 
\begin{eqnarray}
\nonumber
I \simeq {\mathcal P}\int_{k_{i-n_0}}^{k_{i+n_0}} g(q) \dd q  {1- k_i q \over k_i-q}
+{1 \over L} \sum_{j=1}^{i-n_0-1}  {1- k_i k_j \over k_i-k_j}
\\\nonumber
+{1 \over L} \sum_{j=i+n_0+1}^{N}  {1- k_i k_j \over k_i-k_j}
+{1 \over 2 L}  \left[
{1- k_i k_{i-n_0-1} \over k_i-k_{i-n_0-1}}
\right.\\\nonumber\left.+{1- k_i k_{i+n_0+1} \over k_i-k_{i+n_0+1}}
-{1- k_i k_{1} \over k_i-k_{1}}
-{1- k_i k_{N} \over k_i-k_{N}} \right]
\\ \nonumber
+ \int_{-\theta}^{k_1} g(q) \dd q  {1- k_i q \over k_i-q}
+ \int_{k_N}^{\theta} g(q) \dd q  {1- k_i q \over k_i-q}
\end{eqnarray}
which can be rewritten as
\begin{eqnarray}
\nonumber
I \simeq {\mathcal P}\int_{k_{i-n_0}}^{k_{i+n_0}} g(q) \dd q  {1- k_i q
\over k_i-q
}
+{1 \over L} \sum_{j=1}^{i-n_0-1}  {1- k_i k_j \over k_i-k_j}\\\nonumber
+{1 \over L} \sum_{j=i+n_0+1}^{N}  {1- k_i k_j \over k_i-k_j}\\\nonumber
+{1 \over 2 L}  \left[  {1- k_i k_{i+n_0+1} \over k_i-k_{i+n_0+1}}
+ {1- k_i k_{i-n_0-1} \over k_i-k_{i-n_0-1}}
 \right]
\\\label{m1} +{1- k_i k_1 \over k_i-k_1}
\left[ -{1 \over 2 L} +
 \int_{-\theta}^{k_1} g(q) \dd q  \right]\\\nonumber
+{1- k_i k_{N} \over k_i-k_{N}} \left[ -{1 \over 2 L} +\int_{k_N}^{\theta}
g(q) \dd q  \right] 
\end{eqnarray}

From (\ref{kin}) 
 one can show that
\begin{equation}
 {\mathcal P}\int_{k_{i-n_0}}^{k_{i+n_0}} g(q) \dd q  {1- k_i q \over
k_i-q} \simeq {2 k_i n_0 \over L} - {k_i n_0 (1-k_i^2) g'(k_i) \over L
g(k_i)} 
\label{m2} \end{equation}
and that 
\begin{equation}
{1- k_i k_{i+n_0} \over k_i-k_{i+n_0}}
+ {1- k_i k_{i-n_0-1} \over k_i-k_{i-n_0-1}} \simeq   2 k_i -
(1-k_i^2)  {g'(k_i) \over g(k_i)} +  {\mathcal O} \left({1 \over L} \right)
\label{m3} \end{equation}
Lastly because one expects the symmetry $k_j=-k_{N+1-j}$ and because $L
\int_{-\theta}^\theta g(q) dq = N$, one gets that $\alpha =1/2$ in (\ref{jalpha})
and therefore the last two terms of (\ref{m1}) vanish.
\\
Then using (\ref{m2},\ref{m3}) into (\ref{m1}),
one gets that
\begin{equation}\begin{split}
{1 \over L} \sum_{j=1}^{i-n_0-1}  {1- k_i k_j \over k_i-k_j}
+{1 \over L} \sum_{j=i+n_0+1}^{N}  {1- k_i k_j \over k_i-k_j}
 \\\simeq I - {1 \over L} \left( 2 k_i -  (1-k_i^2) {g'(k_i) \over
g(k_i)} \right) \left( n_0 +{1 \over 2} \right)
\label{sum2bis}
\end{split}\end{equation}
where the integral $I$ is defined in (\ref{Idef}).
Lastly  using the fact that $g(k)=g(-k)$,   one  can rewrite the integral $I$ 
in (\ref{Idef}) as 
\begin{equation}
I={\mathcal P}\int_{-\theta}^{\theta} g(k') \dd k'  {1- k'^2  \over k_i-k'}
\end{equation}
so that (\ref{sum2bis}) becomes 
\begin{equation}\begin{split}
{1 \over L} \sum_{j=1}^{i-n_0-1}  {1- k_i k_j \over k_i-k_j}
+{1 \over L} \sum_{j=i+n_0+1}^{N}  {1- k_i k_j \over k_i-k_j}
 \\\simeq  {\mathcal P}\int_{-a}^{a} g(k') \dd k'  {1- k'^2  \over k_i-k'} \\- {1 \over L} \left( 2 k_i -  (1-k_i^2) {g'(k_i) \over
g(k_i)} \right) \left( n_0 +{1 \over 2} \right)
\label{sum2}
\end{split}\end{equation}
Note that (\ref{jalpha})  is not accurate for $i$ close to $1$ or $N$, i.e. near the singularities of $g(k)$. A more detailed analysis of these two neighborhoods would only contribute to higher orders in the $1/L$ expansion \cite{GK}.

\begin{center}
APPENDIX C: SOLUTION OF THE  AIRFOIL  EQUATION (\ref{airfoil-solution})
\end{center}
In this appendix we  show, in the spirit of  \cite{tricomi}, that the solution $\phi(x)$ of 
\begin{equation}
\label{C1}
  f(x) =  {\cal P}
\int_{-1}^1 \dd y\, \frac{\phi(y)}{y-x}
\end{equation}
is 
\begin{equation}
\label{C2}
  \phi(x) = \frac C{\sqrt{1-x^2}}-\frac{1}{\pi^2} {\cal P}\int_{-1}^1 \dd y\, 
  \sqrt{\frac{1-y^2}{1-x^2}}  \frac{f(y)}{y-x}
\end{equation}
This solution is used  to obtain (\ref{phi-sol})  as the solution of
(\ref{airfoil-solution}).

Let us choose
\begin{equation}
\label{C5}
 \phi(x)= {\sqrt{1-x^2} \over x- \alpha} 
\end{equation}
Then  for $x \notin [-1,1]$ 
and $\alpha \notin [-1,1] $
one  can see using (\ref{C4})
\begin{equation}
\label{C6}
  \int_{-1}^1 \dd y\, \frac{\phi(y)}{y-x}
= \pi \left[{\sqrt{\alpha^2-1} \over \alpha -x}  
- {\sqrt{x^2-1} \over \alpha -x} -1  \right]
\end{equation}
and therefore
\begin{equation}
\label{C7}
f(x) =
{\cal P} \int_{-1}^1 \dd y\, \frac{\phi(y)}{y-x} =
\pi \left[ 
 {\sqrt{\alpha^2-1} \over \alpha -x} -1  \right]
\end{equation}
Now the following integral of this function $f(x)$ can be computed (using (\ref{C3},\ref{C4})   for 
  for $x \notin [-1,1]$ 
\begin{equation}\begin{split}
\label{C8}
-\frac{1}{\pi^2} 
 \int_{-1}^1 \dd y\,
 \frac {\sqrt{1-y^2}}{y-x} f(y) =  \sqrt{\alpha^2 -1} \left( {\sqrt{\alpha^2 -1} \over \alpha - x} \right.\\\left.-  {\sqrt{x^2 -1} \over \alpha - x} - 1 \right) + \sqrt{x^2 -1} - x
\end{split}\end{equation}
so that
\begin{equation}\begin{split}
\label{C9}
-\frac{1}{\pi^2} 
{\cal P} \int_{-1}^1 \dd y\,
 \frac {\sqrt{1-y^2}}{y-x} f(y) &=  {\alpha^2 -1
 \over \alpha - x} -    \sqrt{\alpha^2 -1}   
- x \\&= \alpha - \sqrt{\alpha^2 -1} - {1 - x^2 \over \alpha -x} 
\end{split}\end{equation}
Comparing with (\ref{C5}) we see that
\begin{equation}\begin{split}
\label{C10}
-\frac{1}{\pi^2} 
   {\cal P} \int_{-1}^1 \dd y\,
  \sqrt{\frac{1-y^2}{1-x^2}}  \frac{f(y)}{y-x}
&= {\alpha - \sqrt{\alpha^2 -1} \over \sqrt{1- x^2}}
+ {\sqrt{1- x^2} \over x- \alpha} 
\\&  =  {\alpha - \sqrt{\alpha^2 -1} \over \sqrt{1- x^2}} + \phi(x)
\end{split}\end{equation}
Therefore (\ref{C2}) is the solution of (\ref{C1})  with a constant $C$ which depends  through $\alpha$ on $\phi(x)$ when one chooses (\ref{C5}) for $\phi(x)$.

As the inversion formula (\ref{C2}) is valid for arbitrary $\alpha$, it
would also be valid when $f(x)$ is  any polynomial  in $x$, and as the
polynomials are dense in the set of continuous functions on $(-1,1)$, one
can consider that (\ref{C1},\ref{C2}) are valid for "arbitrary functions"
$f(x)$.

\begin{center}
APPENDIX D: USEFUL INTEGRALS
\end{center}
In this appendix we list a few integrals which are used in various places
of the paper.

First for $x \notin [-1,1]$  one has
\begin{equation}
\label{C3}
{1 \over \pi}\int_{-1}^1 {\sqrt{1-y^2} \over y -x} dy  = \sqrt{x^2-1}-x 
\end{equation}
so that \begin{equation}
\label{C3bis}
{1 \over \pi} {\cal P} \int_{-1}^1 {\sqrt{1-y^2} \over y -x} dy  = -x
\end{equation}

As a consequence of (\ref{C4})  one  has for $x \notin [-1,1]$ and $\alpha \notin [-1,1] $
\begin{equation}
\label{C4}
{1 \over \pi}\int_{-1}^1 {\sqrt{1-y^2} \over y -x} {dy \over y - \alpha}= 
{\sqrt{\alpha^2-1} \over \alpha -x}
-{\sqrt{x^2-1} \over \alpha -x} -1 
\end{equation}
and thus for  $x \in [-1,1]$ and $\alpha \notin [-1,1] $
\begin{equation}
\label{C4bis}
{1 \over \pi} {\cal P} \int_{-1}^1 {\sqrt{1-y^2} \over y -x} {dy \over y - \alpha}= 
{\sqrt{\alpha^2-1} \over \alpha -x}
 -1 
\end{equation}

One can also show that 
\begin{equation}
\label{B3}
 \int_{-1}^1 {dx \over \sqrt{1-x^2}} = \pi 
\end{equation}
and that for $y \notin [-1,1]$
\begin{equation}
\label{B4}
 \int_{-1}^1 {dx \over \sqrt{1-x^2}}  {1 \over y-x}=  {\pi  \over \sqrt{y^2-1}}
\end{equation}
As a consequence of (\ref{C4bis},\ref{B4}),  one has 
\begin{equation}
\label{B5}
 \int_{-1}^1 {dx \over \sqrt{1-x^2}}  \ {\cal P} \int_{-1}^1 {dy \over y-x} F(y) =  0 
\end{equation}
for an arbitrary function $F(y)$ as it is valid for any polynomial .


For $\theta<1$ one can show using (\ref{B4}) that 
\begin{equation}
\label{B6}
 \int_{-1}^1 {dx \over  (1- \theta^2 x^2) \sqrt{1-x^2}}  =  {\pi  \over
\sqrt{1-\theta^2}} 
\end{equation}
one can also show
\begin{equation}
\label{B7}
 \int_{-1}^1 { 
\sqrt{1-x^2}
\over  (1- \theta^2 x^2) 
}  dx =  \pi {1 - \sqrt{1-\theta^2} \over
\theta^2} 
\end{equation}
and that
\begin{equation}
\label{B20bis}
\int_{-1}^1 {y^2 \sqrt{1-y^2} \over 1 - \theta^2 y^2} dy  = \pi \left( {1 \over \theta^4} - {1 \over 2 \theta^2} - {\sqrt{1-\theta^2} \over \theta^4} \right)
\end{equation}
and  for $y \notin[-1,1]$
\begin{equation}\begin{split}
\label{B7bis}
 \int_{-1}^1 {1 \over  (1-\theta^2 x^2) \sqrt{1-x^2}} \  {dy \over y-x}
=&  {\pi \over (1- \theta^2 y^2) \sqrt{y^2-1} } \\& -   { \pi \theta^2 y \over (1-\theta^2 y^2) \sqrt{1 - \theta^2}} 
\end{split}\end{equation}
and therefore for any function $F(y)$ 
\begin{equation}\begin{split}
\label{B8}
 \int_{-1}^1 {dx \over (1-\theta^2 x^2) \sqrt{1-x^2}}
 \ {\cal P} \int_{-1}^1 {dy \over y-x} F(y)= \\  -  {\pi \theta^2 \over
\sqrt{1-\theta^2}} \int_{-1}^1 {y
F(y) \over 1 - \theta^2 y^2 } dy  \ .
\end{split}\end{equation}

\begin{center}
APPENDIX E: BETHE ANSATZ CALCULATION FOR THE CURRENT LARGE DEVIATION FUNCTION $\psi_Q(s)$
\end{center}

This appendix describes  how a Bethe ansatz calculation of $\psi_Q(s)$
similar to the one  conducted for  $\psi_K$ can be implemented. The
operator $\mathbb{W}_Q$ whose largest eigenvalue is $\psi_Q$ 
reads, in the spin language already used in (\ref{defHQ-defHK}),
\begin{equation}\label{evolQs}
 \mathbb{W}_Q(s ) =  \sum_{i=1}^L  \left[
   \frac{\sigma_i^{z}\sigma_{i+1}^{z}-1}{2}
  +  \ee^{-s}   \sigma_i^+\sigma_{i+1}^- 
  + \ee^{s}\sigma_i^-\sigma_{i+1}^+
 \right]
\end{equation}
The Bethe ansatz equation analogous to (\ref{eqn:Bethe_eqs}) take the form
(\ref{bethe-Q})
\begin{equation}\label{BetheqforQ}
\zeta_i^L=\prod_{\substack{ j=1\\ j\neq i}}^{N} 
\left[-\frac{1-2  \ee^{-s} \zeta_i+\ee^{-2s}\zeta_i \zeta_j}
            {1-2 \ee^{-s} \zeta_j+\ee^{-2s}\zeta_i \zeta_j}\right]
\end{equation}
In terms of the $\zeta_j$'s, we have that
\begin{equation}
 \psi_Q(s)=-2N+e^{-s}  \big[\zeta_1+\ldots+\zeta_N\big]
                +e^{s}\left[\frac{1}{\zeta_1}+\ldots+\frac{1}{\zeta_N}\right]
\end{equation}
Kim~\cite{kim} has studied the spectrum of ${\cal H}=-\mathbb{W}_Q/(\cosh s/2)$ by means of a Bethe ansatz calculation: in the notations of his equation (1), the parameters $\tilde{\Delta}$ and $S$ are given by 
\begin{equation}
\tilde\Delta=\frac 1{\cosh s} \:,\quad  S=\tanh s 
\end{equation}
but unfortunately his results do not apply to our particular case, which turns out to correspond to a critical point of the related six-vertex model. The defining parameters of the latter, denoted by $\Delta$, $H$ and $\nu$, are related to Kim's by $\tilde\Delta=\Delta/\cosh(2H)$,
$S=\tanh(2H)$, $\Delta=\cosh\nu$. Thus, in terms of our original parameters, we get that 
\begin{equation}
  \Delta = 1
  \:,\quad 2H=s
  \:,\quad \nu=0
\end{equation}
a limiting case explicitly excluded by Kim which lies at the critical point of the six-vertex model.\\

We choose to write that $\zeta_j=\ee^{-is(k_j+2i\rho)}$. The two main
differences with the calculation of $\psi_K$  is that  the $\zeta_j$'s dependence in $s$ is different. We have also shifted them by $2i\rho$ for convenience. Just
as was the case previously, the $k_j$'s will be densely distributed on a
connected curve $\cal C$ of the complex plane that is invariant upon
complex conjugation. Given that the equations for the $\zeta_j$'s are invariant under complex conjugation, we expect the contour $\cal C$ to be symmetric with respect to the vertical axis in the complex $k$ plane. We shall denote the end points of $\cal C$ by $-\theta^*$ and $\theta$.\\ 

Given that (\ref{BetheqforQ}) becomes
\begin{equation}\begin{split}
-i(k_i+i2\rho)&=\frac{1}{L}\sum_{j=1,j\neq i}^{N}U(k_i,k_j),\; {\rm where} \\  U(k_i,k_j)&=\frac{1}{s}\ln\left[-\frac{1-2  \ee^{-s} \zeta_i+\ee^{-2s}\zeta_i \zeta_j}
            {1-2 \ee^{-s} \zeta_j+\ee^{-2s}\zeta_i \zeta_j}\right]
\end{split}\end{equation}
for $|i-j|\gg1$ we expect that
\begin{equation}\label{BQ1}
U(k_i,k_j)=\frac{2i (k_i+i\alpha) (k_j+i\alpha)}{k_i-k_j},\;\;\alpha=2\rho-1
\end{equation}
while for $i-j$ of order 1, $s$ will be over order $1/L$ and $k_i-k_j$ as
well. We define $g(k)$ as the root density along contour $\cal C$, so that
$$L \int_{k_i}^{k_j} g(k) dk = j-i$$
(note that $g(k)$ is in general complex but along the contour $g(k) dk$ is real).
If $k_j$ and $k_i$ are $n$ roots apart, we have that

$k_j-k_i=\frac{n}{g(k_i)L}-\frac{n^2 g'(k_i)}{2g(k_i)^3L^2}+\ldots$. Expanding $U$ at fixed $sL$ in powers of $L^{-1}$ leads to
\begin{equation}\label{BQ2}\begin{split}
U(k_i,k_j)=\frac{1}{s}\ln\frac{n-ig(k_i)(k_i+i\alpha)^2 sL}{n+ig(k_i)(k_i+i\alpha)^2 sL}\qquad\qquad\\
-{i}(k_i+i\alpha)\left(2+\frac{g'(k_i)(k_i+i\alpha)}{g(k_i)}\right)\frac{n^2}{n^2+[g(k_i) (k_i+i\alpha)^2 (sL)]^2}
\end{split}\end{equation}
Equations (\ref{BQ1}) and (\ref{BQ2}) play a role analogous to (\ref{U1}) and (\ref{UU1}) in the study of $K$. After using the methods of appendices B and C we arrive at the following equation for $g$ which we express in terms of $\phi(x)=(\theta x+i\alpha)^2 g(\theta x)$ and $r=\theta^*/\theta$:
\begin{equation}\begin{split}\label{defphi}
\theta\left(x+i\frac{\alpha+1}{\theta}\right)=2{\cal P}\!\!\int_{-r}^1\!\!\!\!\dd y \frac{\phi(y)-(y-x)(y+i\frac{\alpha}{\theta})^{-1}\phi(y)}{y-x}\\
-\frac{\theta}{L}\left(x+i\frac{\alpha}{\theta}\right)^2\frac{\phi'(x)}{\phi(x)}[\pi \phi(x) (sL)]\coth[\pi \phi(x) (sL)]\end{split}\end{equation}
Let us denote $\phi_0(x)$ the solution of the above equation, in the $L \to \infty$ limit
\begin{equation}\begin{split}
\theta x/2+h=&{\cal P}\int_{-r}^{1}{\dd y}\frac{\phi_0(y)}{y-x}\end{split}
\label{eq130}
\end{equation}
where $h=i(\alpha+1)/2+\int\dd y\phi_0(y)(y+i\alpha/\theta)^{-1}$ is a density-dependent constant to be determined. The general solution of (\ref{eq130}) can be written (see (\ref{C1},\ref{C2})) as 
\begin{equation}\label{solphi0}\begin{split}
\phi_0(x)=&-\frac{C}{\sqrt{(1-x)(r+x)}}-\frac{\theta(r+1)^2}{16\pi\sqrt{(1-x)(r+x
)}}
\\&+\frac{\theta x^2}{2\pi\sqrt{(1-x)(r+x)}}+\frac{2 h (r-1+x)}{2\pi\sqrt{(1-x)(r+x
)}}\\&+\frac{\theta x(r-1)}{4\pi\sqrt{(1-x)(r+x)}}
\end{split}\end{equation}
The four unknowns $C$, $\theta$, $r$ and $h$ are determined by requiring that $\phi_0$ remains finite as $x\to 1$ and as  $x\to -r$, and by noting that by definition
\begin{equation}\label{appE1}
\int_{-r}^{+1}\dd x\frac{\phi_0(x)}{\left(x+i\frac{\alpha}{\theta}\right)^2}=\theta \rho
\end{equation}
while $\phi_0$ must verify the self-consistency equation $h=i(\alpha+1)/2+\int\dd y\phi_0(y)(y+i\alpha/\theta)^{-1}$.  After explicitly evaluating the latter integral and that appearing in (\ref{appE1}), we arrive at $r=1$, $h=0$ and $4\pi C=\theta=2\sqrt{\rho(1-\rho)}$, which leads to $\phi_0(x)=-\theta\frac{\sqrt{1-x^2}}{2\pi}$. Up to a sign, this is exactly the same function as that found in the study of $K$, and this is the same end point $\theta=2\sqrt{\rho(1-\rho)}$ for the contour on which the $k_j$'s lie.\\

We may now simplify (\ref{defphi}) into
\begin{equation}\begin{split}
\theta\left(x+i\frac{\alpha+1}{\theta}\right)=2{\cal P}\!\!\int\!\!{\dd y}\frac{\phi(y)-(y-x)(y+i\alpha/\theta)^{-1}\phi(y)}{y-x}\\
+\frac{1}{\theta L}\frac{x(x+i\alpha/\theta)^2}{1-x^2}\left([\theta\sqrt{1-x^2} (sL)/2]\coth[\theta\sqrt{1-x^2}(sL)/2]\right)
\end{split}\end{equation}
whose solution reads $\phi(x)=\phi_0(x)+\delta\phi(x)$,
\begin{equation}\begin{split}
\delta\phi(x)=&-\frac{\delta C}{\sqrt{1-x^2}}+\frac{2\delta h x}{2\pi\sqrt{1-x^2}}\\
&-\frac{1}{\pi^2}\frac{1}{\sqrt{1-x^2}}{\cal P}\int\dd y\frac{\sqrt{1-y^2}}{y-x}\delta F(y)
\end{split}\end{equation}
We have denoted by $\delta F(x)$ the function
\begin{equation}\begin{split}
\delta F(x)=-\frac{\theta}{2L}\frac{x(x+i\alpha/\theta)^2}{1-x^2}\qquad\qquad\qquad\qquad\\\times\left([\theta\sqrt{1-x^2} (sL)/2]\coth[\theta\sqrt{1-x^2}(sL)/2]\right)
\\
=-\frac{\theta}{2L}\frac{x(x+i\alpha/\theta)^2}{1-x^2}\left[\sum_{p\geq 2}\frac{B_p}{p!}(\theta sL)^p(1-x^2)^{p/2}+1\right]
\end{split}
\end{equation}
The new constants $\delta C$ and $\delta h$ are determined by $\int\frac{\delta\phi}{(x+i\alpha/\theta)^2}=0$ and 
$\delta h=\int\frac{\delta\phi}{(x+i\alpha/\theta)}$. After performing explicit integrations along the lines of appendix D, we obtain the final result through the following equality
\begin{equation}\begin{split}\label{alm}
\psi_Q(s)/L=-s^2\theta\int\dd x\phi(x)=\frac{\theta^2}{4}s^2+s^2\delta C\theta\pi\\+s^2 \theta\frac{1}{\pi^2}\int\dd x\frac{1}{\sqrt{1-x^2}}{\cal P}\int\dd y\frac{\sqrt{1-y^2}}{y-x}\delta F(y)
\end{split}\end{equation}
where 
\begin{equation}\begin{split}
\delta C\theta\pi=\frac{\theta^2}{2\pi L}\int_{-1}^1\dd x\; x^2\left[\sum_{p\geq 2}\frac{B_p}{p!}(\theta sL)^p(1-x^2)^{\frac{p-1}{2}}\right.\\\left.+\frac{1}{\sqrt{1-x^2}}\right]\\
=\frac{1}{L^3 s^2}{\cal F}(-L^2s^2\theta^2/8)+\frac{\theta^2}{4L}
\end{split}\end{equation}
After noting that, as before, we have
\begin{equation}
\frac{1}{\pi^2}\int\dd x\frac{1}{\sqrt{1-x^2}}{\cal P}\int\dd y\frac{\sqrt{1-y^2}}{y-x}\delta F(y)=0
\end{equation} 
it only remains to substitute  the value of $\delta C$ into (\ref{alm}). This allows us to conclude that
\begin{equation}
\psi_Q(s)=\frac{\theta^2}{4}s^2(L+1)+L^{-2}{\cal F}(-L^2s^2\theta^2/8)
\end{equation}
which is the announced result of (\ref{finalpsiQ}).

\end{document}